\documentclass[twocolumn,usenatbib]{mn2e}

\usepackage{amsmath,amssymb,bm,color,float,graphicx,dcolumn,Macro,mathrsfs,subfigure,tabularx}

\def\T{\Theta}
\def\bX{\ol{X}}
\def\bY{\ol{Y}}
\def\tT{\widetilde{\T}}
\def\tE{\widetilde{E}}
\def\tB{\widetilde{B}}
\def\tX{\widetilde{X}}
\def\tY{\widetilde{Y}}
\def\tXi{\widetilde{\Xi}}
\def\hT{\widehat{\T}}
\def\hE{\widehat{E}}
\def\hB{\widehat{B}}
\def\hX{\widehat{X}}
\def\hY{\widehat{Y}}
\def\hXi{\widehat{\Xi}}
\def\hPsi{\widehat{\Psi}}
\def\hCXX{\widehat{C}^{XX}}
\def\hCXY{\widehat{C}^{XY}}
\def\hCYY{\widehat{C}^{YY}}
\def\tCTT{\widetilde{C}^{\T\T}}
\def\tCTE{\widetilde{C}^{\T E}}
\def\tCEE{\widetilde{C}^{EE}}
\def\tCBB{\widetilde{C}^{BB}}
\def\Beam{\mC{B}}
\def\estg{\widehat{\grad}}

\def\estx{\widehat{x}}
\def\esty{\widehat{y}}
\def\barx{\ol{x}}

\def\ol{\overline}
\def\bl{\bm{\ell}}
\def\bL{\bm{L}}

\def\hatn{\hat{\bm{n}}}
\def\polangle{\alpha}
\def\hpolangle{\widehat{\alpha}}
\def\pols{\psi}

\title{Bias-Hardened CMB Lensing with Polarization} 

\author[T. Namikawa et al.]
{
	Toshiya Namikawa$^1$\thanks{E-mail: namikawa@yukawa.kyoto-u.ac.jp} 
	and Ryuichi Takahashi$^2$ \\ 
	$^1$Yukawa Institute for Theoretical Physics, Kyoto University, Kyoto 606-8502, Japan \\ 
	$^2$Faculty of Science and Technology, Hirosaki University, 3 bunkyo-cho, Hirosaki, Aomori, 036-8561, Japan
}

\begin{document}

\maketitle

\begin{abstract}
Polarization data will soon provide the best avenue for measurements of the CMB lensing potential, 
although it is potentially sensitive to several instrumental effects including beam asymmetry,
polarization angle uncertainties, sky coverage, as well as analysis choices such as masking.
We derive ``bias-hardened'' lensing estimators to mitigate these effects, at the expense of 
somewhat larger reconstruction noise, and test them numerically on simulated data.
We find that the mean-field bias from masking is significant for the $EE$ quadratic lensing estimator,
however the bias-hardened estimator combined with filtering techniques can mitigate the mean field. 
On the other hand, the $EB$ estimator does not significantly suffer from the mean-field from 
the point source masking and survey window function. 
The contamination from beam asymmetry and polarization angle uncertainties, however, can generate 
mean-field biases for the $EB$ estimator. These can also be mitigated using bias-hardened estimators, 
with at most a factor of $\sim 3$ degradation of noise level compared to the conventional approach. 
\end{abstract} 


\section{Introduction} \label{sec.1}

On arcminute scales, the CMB temperature and polarization anisotropies are distorted by gravitational 
lensing. For the past several years, CMB observations have been used to make increasingly precise 
measurements of this effect, both with cross-correlations between CMB and large-scale structure 
(\citealt{Smith07,Hirata:2008cb,Bleem:2012gm,Sherwin:2012mr,Ade:2013aro,Holder:2013hqu,
Geach:2013zwa,Hanson:2013daa}) as well as CMB maps alone (\citealt{Das:2011ak,vanEngelen:2012va,
Das:2013zf,Ade:2013tyw,Hanson:2013daa}). 

These lensing measurements are already being used to constrain cosmology 
(e.g., \citealt{Sherwin:2011gv,vanEngelen:2012va,Ade:2013zuv,Battye:2013xqa,
Namikawa:2013wda,Wilkinson:2013kia}), future measurements are expected to quantify the sum of 
neutrino masses (e.g., \citealt{Namikawa:2010re,Joudaki:2011nw,Abazajian:2013oma} and refs. therein),
and provide even tighter constraints on cosmic strings (e.g.,\citealt{Namikawa:2011cs,Yamauchi:2012bc,
Yamauchi:2013fra}), primordial non-Gaussianity (e.g., \citealt{Jeong:2009wi,Takeuchi:2011ej}), 
and other fundamental physics. Lensing potential estimates should also be important for 
{\it delensing} \citep{Knox:2002pe,Kesden:2002ku} to detect inflationary gravitational waves at 
$>10$ if the tensor-to-scalar ratio is less than $r\sim0.01$.

Given an observed CMB, estimators to reconstruct the lensing potential have been derived by several 
authors (e.g., \citealt{Zaldarriaga:1998te,Seljak:1998aq,Hu:2001kj,Okamoto:2003zw,Hirata:2003ka,
Namikawa:2011cs}). These estimators all utilize the fact that a fixed lensing potential introduces 
statistical anisotropy into the observed CMB, in the form of a correlation between the CMB 
temperature/polarization anisotropies and their gradients. With a large number of observed CMB modes, 
this correlation may be used to form estimates of the lensing potential. The power spectrum of the 
lensing potential, which is of more interest for cosmological parameter constraints,
can then be estimated from the power spectrum of these estimates (which probes the non-Gaussian 
4-point function of the lensed CMB). For CMB observations with noise levels below 
$5\mu K\! \cdot\!{\rm arcmin}$, $B$-mode polarization is a particularly powerful probe of lensing 
as it is believed to be dominated by the lensing contribution on scales $\gtrsim 100$.

For realistic CMB observations, there are so called {\it mean-field biases} for the standard 
minimum-variance quadratic lensing estimators due to non-lensing sources of statistical 
anisotropy such as masking, inhomogeneous map noise, beam asymmetry, or spatially-varying errors in 
the detector polarization angles. With perfect statistical understanding of the unlensed CMB and the 
instrument used to observe it, these biases may be corrected for, however given imperfections in our 
understanding of these quantities it can be useful to design estimators which are less sensitive to 
them.

Approaches have been proposed in the literature to mitigate some of the mean-field biases. The mean 
fields from masking in temperature, for example, have been studied by several authors, with approaches 
including simply avoiding mask boundaries (\citealt{Hirata:2008cb,Carvalho:2011gx}), 
or using inpainting/apodization (\citealt{Perotto:2009tv,Plaszczynski:2012ej,BenoitLevy:2013bc}) to 
smooth them. These techniques could also be utilized for polarization, in conjunction with 
``pure'' estimators for $E$ and $B$ modes \citep{Smith:2006hv,Smith:2006vq}. 
For temperature case, the mean-field bias from inhomogeneous map noise is also studied in 
\citet{Hanson:2009dr}. 

In this paper, we extend the bias-hardened estimators proposed in our previous work 
\citep{Namikawa:2012pe} to the case of lensing reconstruction with polarization, constructing lensing 
estimators which can have significantly smaller mean-field biases than the standard minimum-variance 
estimators, with minimal loss of signal-to-noise. 

This paper is organized as follows.
In Sec.~\ref{sec.2}, we briefly summarize quadratic estimators for the lensing potential using
CMB temperature and polarization.
In Sec.~\ref{sec.3}, we discuss several possible mean-field biases which must be corrected for, and 
then construct corresponding bias-hardened estimators. 
In Sec.~\ref{sec.4}, we demonstrate the usefulness of bias-hardened estimators using numerical 
simulations. Sec.~\ref{sec.5} summarizes our results.

Finally, we note that when estimating the power spectrum of the lensing potential, there is an
additional worrisome bias, the \textit{reconstruction noise} bias, which must be accounted for.
This bias is analogous to shape-noise in galaxy weak lensing measurements; in principle it 
can be avoided by forming independent lensing estimates from subsets of the observed sky modes,
for example using an odd/even parity split \citep{Hu:2001fa} or the in/out Fourier split 
\citep{Sherwin:2010ge}. 
There is, however, usually a substantial loss of signal-to-noise associated with such splits. 
This bias is less of an issue in polarization than in temperature, because it falls with the 
instrumental noise level for estimates which utilize $B$-mode polarization, and can also be avoided 
using cross-spectra between different lensing estimates.
In appendix \ref{app:BHE} we present a polarized derivation of the optimal trispectrum-estimator
approach to correcting for this bias, also extending our discussion of this approach for the 
temperature case in \citep{Namikawa:2012pe}.

\section{Quadratic lensing reconstruction from CMB maps} \label{sec.2}

\begin{table}
\renewcommand{\arraystretch}{1.8}
\bc
\caption{
The weight functions for lensing potentials, $f_{\bl,\bL}^{x,(XY)}$. Note that $\bL'=\bl-\bL$. 
}
\label{table:weight}
\begin{tabular}{cc} \hline 
\multicolumn{2}{c}{Lensing} \\ \hline 
$\T\T$ & $c^{ab}_x\{\l_a L_b \tCTT_L+\l_a L'_b \tCTT_{L'} \}$ \\
$\T E$ & $c^{ab}_x\{\l_a L_b \tCTE_L\cos 2\varphi_{\bL,\bL'}+\l_aL'_b\tCTE_{L'}\}$ \\
$\T B$ & $c^{ab}_x\{\l_a L_b\tCTE_L\sin 2\varphi_{\bL,\bL'}$ \\ 
$EE$ & $c^{ab}_x\{\l_a L_b \tCEE_L+\l_a L'_b \tCEE_{L'} \} \cos 2\varphi_{\bL,\bL'}$ \\
$EB$ & $c^{ab}_x\{\l_a L_b \tCEE_L+\l_a L'_b \tCBB_{L'} \} \sin 2\varphi_{\bL,\bL'}$ \\ 
$BB$ & $c^{ab}_x\{\l_a L_b \tCBB_L+\l_a L'_b \tCBB_{L'} \} \cos 2\varphi_{\bL,\bL'}$ \\ \hline
\end{tabular}
\ec 
\end{table}

\subsection{Lensing effect on CMB anisotropies} 

The distortion effect of lensing on the primary temperature and polarization anisotropies is expressed 
by a remapping of the primary anisotropies. Denoting the primary CMB anisotropies at position 
$\hatn=(\theta,\varphi)$ on the last scattering surface as $\Xi^{(s)}(\hatn)$, where $s=0$ denotes the 
temperature, $\Xi^{(0)}=\T$, while $s=\pm 2$ are the spin-$2$ combination of the Stokes parameter, 
$\Xi^{(\pm 2)}=Q\pm\iu U\equiv P^\pm$, the lensed anisotropies in a direction $\hatn$, are given 
by (e.g., \citealt{Lewis:2006fu})
\al{
	\tXi^{(s)}(\hatn) &= \Xi^{(s)}(\hatn+\bm{d}(\hatn)) 
	\notag \\ 
		&= \Xi^{(s)}(\hatn) + d^a(\hatn)\pd_a \Xi^{(s)}(\hatn) + \mC{O}(|\bm{d}|^2)
	\,. \label{eq:remap}
}
The two-dimensional vector, $d^a(\hatn)$ $(a=\theta,\varphi)$, is the deflection angle, and, 
in terms of parity symmetry, we can decompose it into two terms, known as gradient (even parity) 
and curl (odd parity) modes (e.g., \citealt{Hirata:2003ka,Cooray:2005hm,Namikawa:2011cs}): 
\al{
	d^a(\hatn) = \pd^a\grad (\hatn) + \epsilon^{ab}\pd_b \curl (\hatn) 
		= \sum_{x=\grad,\curl} c^{ab}_x\pd_b x(\hatn) 
	\,, \label{eq:deflection}
}
where the symbol, $c^{ab}_\grad$, is the Kronecker delta and $c^{ab}_\curl=\epsilon^{ab}$ is the 
two-dimensional Levi-Chivita symbol. 

\subsection{Estimator for lensing fields} 

The temperature anisotropies are distorted by lensing as (e.g., \citealt{Hu:2001kj}) 
\footnote{
Our definitions of Fourier transform and its inverse for arbitrary quantity $X(\hatn)$ on a map are 
\al{
	X_{\bl} &= \FT{\hatn}{\bl}{f} X(\hatn) \,, \\ 
	X(\hatn) &= \FT{\bl}{\hatn}{b} X_{\bl} \,. 
}
These are the same as \citet{Hu:2001kj} but different from, e.g., \citet{Lewis:2006fu}. 
}
\al{
	\tT_{\bl} &= \T_{\bl} - \sum_{x=\grad,\curl}\Int{2}{\bL}{(2\pi)^2}c_x^{ab}L_a(_b-L_b)\grad_{\bL}\T_{\bl-\bL} 
	\,. 
}
On the other hand, for polarization we usually use the rotationally invariant combination, 
i.e., the E and B mode polarizations, instead of the spin-2 quantity (e.g., \citealt{Hu:2001kj}): 
\al{
	E_{\bl} \pm i B_{\bl} = - \FT{\hatn}{\bl}{f} P^{\pm}(\hatn) \E^{\mp 2\iu\varphi_{\bl}}
	\,, 
}
where $\varphi_{\bl}$ is the angle of $\bl$ measured from the $x$-axis. 
With deflection angle given in \eq{eq:deflection} , the lensed E and B modes are 
given by (e.g., \citealt{Hu:2001kj,Cooray:2005hm,Namikawa:2011cs})
\al{
	\tE_{\bl} &= E_{\bl} - \Int{2}{\bL}{(2\pi)^2} (L^a L'_a \grad_{\bL} + \epsilon^{ab}L_aL'_b \curl_{\bl}) 
	\notag \\ 
		&\qquad \times (E_{\bL'}\cos 2\varphi_{\bL',\bl} - B_{\bL'}\sin 2\varphi_{\bL',\bl}) 
	\,, \\ 
	\tB_{\bl} &= B_{\bl} - \Int{2}{\bL}{(2\pi)^2} (L^a L'_a \grad_{\bL} + \epsilon^{ab}L_aL'_b \curl_{\bl}) 
	\notag \\ 
		&\qquad \times (B_{\bL'}\cos 2\varphi_{\bL',\bl} + E_{\bL'}\sin 2\varphi_{\bL',\bl}) 
	\,, \label{eq:lensed-EB}
}
with $\bL'=\bl-\bL$ and $\varphi_{\bl_1,\bl_2} \equiv \varphi_{\bl_1}-\varphi_{\bl_2}$. 

Denoting $X$ and $Y$ as $\T$, $E$ or $B$, the off-diagonal covariance includes the gradient and 
curl modes of deflections as 
\al{
	\ave{\tX_{\bL}\tY_{\bL'}}\rom{CMB} &= f_{\bl,\bL}^{x,(XY)} x_{\bl}  \,, \label{Eq:weight}
}
where $\ave{\cdots}\rom{CMB}$ denotes the ensemble average over unlensed $\T$, $E$ or $B$, with a fixed realization of the gradient and curl modes, and we ignore the higher-order terms of lensing fields. The weight functions for gradient and curl modes are summarized in Table \ref{table:weight} \citep{Hu:2001kj,Cooray:2005hm,Namikawa:2011cs}. Note that, to mitigate the higher-order biases \citep{Hanson:2010rp}, the lensed power spectrum is used rather than the unlensed one \citep{Lewis:2011fk,Anderes:2013jw}: With a quadratic combination of $X$ and $Y$ fluctuations, the lensing estimators are then formed as (e.g., \citealt{Hu:2001kj}), 
\al{
	\estx^{(XY)}_{\bl} 
		&= \frac{1}{2}A_\l^{x,(XY)}\Int{2}{\bL}{(2\pi)^2} g_{\bl,\bL}^{x,(XY)} \bX_{\bL}\bY_{\bL'} 
	\,, 
}
where, with the ratio of power spectra, $r^{XY}_L=\hCXY_L/\hCXX_L$, we define 
\footnote{
Note here that the normalization, $A_{\bl}^{x,(XY)}$, is independent of the direction of $\bl$, so 
we write the normalization as $A_\l^{x,(XY)}$. 
}
\al{
	g^{x,(XY)}_{\bl,\bL} 
		&= 2\frac{[f^{x,(XY)}_{\bl,\bL}]^*-r^{XY}_{L}r^{YX}_{L'} [f^{x,(XY)}_{\bl,\bL'}]^*}
			{1-r^{XY}_Lr^{XY}_{L'}r^{YX}_Lr^{YX}_{L'}} 
	\,, \label{Eq:def-gXY} \\ 
	A_\l^{x,(XY)} 
		&= \left\{\Int{2}{\bL}{(2\pi)^2} \frac{g_{\bl,\bL}^{x,(XY)}f_{\bl,\bL}^{x,(XY)}}
			{2\hCXX_{L}\hCYY_{L'}}\right\}^{-1}
	\,. \label{Eq:def-AXY}
}
The inverse-variance filtered Fourier modes are given by 
\al{
	\bX_{\bl} = \frac{\hX_{\bl}}{\hCXX_\l}  \,. 
}
For the cosmic variance case, the estimated power spectrum reduces to the lensed power spectrum. 

\section{Bias-hardened lensing estimators} \label{sec.3} 

\begin{table}
\renewcommand{\arraystretch}{1.8}
\bc
\caption{
The weight functions for masking, $f_{\bl,\bL}^{M,(XY)}$. Note that $\bL'=\bl-\bL$. 
}
\label{table:weight-mask}
\begin{tabular}{cc} \hline 
\multicolumn{2}{c}{Masking} \\ \hline 
$\T\T$ & $-\tCTT_L-\tCTT_{L'}$ \\
$\T E$ & $-\tCTE_L\cos 2\varphi_{\bL,\bL'}-\tCTE_{L'}$ \\
$\T B$ & $-\tCTE_L\sin 2\varphi_{\bL,\bL'}$ \\ 
$EE$ & $-(\tCEE_L+\tCEE_{L'})\cos\varphi_{\bL,\bL'}$ \\ 
$EB$ & $-(\tCEE_L+\tCBB_{L'})\sin\varphi_{\bL,\bL'}$ \\ 
$BB$ & $-(\tCBB_L+\tCBB_{L'})\cos\varphi_{\bL,\bL'}$ \\ \hline
\end{tabular}
\ec 
\end{table}

\begin{table*}
\renewcommand{\arraystretch}{1.8}
\bc
\caption{
Same as Table \ref{table:weight-mask}, but for polarization angle, $f^{\pols^{(n,p)},(XY)}_{\bl,\bL}$. 
For clarity, the exponential is expressed as ``$\exp$". 
}
\label{table:weight-beam}
\begin{tabular}{cc} \hline 
\multicolumn{2}{c}{$p=0$} \\ 
\hline 
$\T\T$ & 
$\big [\Beam^{(\T\T)}_{L,(n,0)}\tCTT_L + \Beam^{(\T E)}_{L,(n,0)}\tCTE_L \big] 
\exp \big(\iu n\varphi_{-\bL,\bl}\big) + (\bL\leftrightarrow \bL') $ \\
$\T E$ & 
$\big [\Beam^{(\T\T)}_{L',(n,0)}\tCTE_{L'} + \Beam^{(\T E)}_{L',(n,0)}\tCEE_{L'}\big]
\exp \big(\iu n\varphi_{-\bL',\bl}\big) $ \\ 
$\T B$ & 
$\Beam^{(\T B)}_{L',(n,0)}\tCBB_{L'}\exp\big(\iu n\varphi_{-\bL',\bl}\big)$ \\
$EE$ & $0$ \\ 
$EB$ & $0$ \\ 
$BB$ & $0$ \\ 
\hline 
\multicolumn{2}{c}{$p=\pm 1$} \\ 
\hline 
$\T\T$ & $0$ \\ 
$\T E$ & 
$\big [\Beam^{(E\T)}_{L,(n,p)}\tCTT_L + \Beam^{(EE)}_{L,(n,p)}\tCTE_L \big ]
\exp \big(\iu n\varphi_{-\bL,\bl}\pm 2\iu\varphi_{-\bL,\bL'}\big)$ \\  
$\T B$ & 
$\big [\Beam^{(B\T)}_{L,(n,p)}\tCTT_L + \Beam^{(BE)}_{L,(n,p)}\tCTE_L \big ]
\exp \big(\iu n\varphi_{-\bL,\bl}\pm 2\iu\varphi_{-\bL,\bL'}\big)$ \\ 
$EE$ & 
$\big [\Beam^{(E\T)}_{L,(n,p)}\tCTE_L + \Beam^{(EE)}_{L,(n,p)}\tCEE_L \big]
\exp \big(\iu n\varphi_{-\bL,\bl}\pm 2\iu\varphi_{-\bL,\bL'}\big) + (\bL\leftrightarrow \bL') $ \\ 
$EB$ & 
$\big [\Beam^{(B\T)}_{L,(n,p)}\tCTE_L + \Beam^{(BE)}_{L,(n,p)}\tCEE_L \big ]
\exp \big(\iu n\varphi_{-\bL,\bl}\pm 2\iu\varphi_{-\bL,\bL'}\big) 
+ \Beam^{(EB)}_{L',(n,p)}\tCBB_{L'}
\exp \big(\iu n\varphi_{-\bL',\bl}\pm 2\iu\varphi_{-\bL',\bL}\big)$ \\ 
$BB$ & 
$\Beam^{(BB)}_{L,(n,p)}\tCBB_L \exp \big(\iu n\varphi_{-\bL,\bl}\pm 2\iu\varphi_{-\bL,\bL'}\big)
+ (\bL\leftrightarrow \bL') $ \\
\hline
\end{tabular}
\ec 
\end{table*}

There are many effects which can generate mode-coupling between observed $\T$, $E$ and $B$ modes, 
leading to mean-field biases for the conventional lensing estimators. In the following, we compute 
the non-lensing statistical anisotropy due to masking, inhomogeneous noise 
(and/or unresolved point sources), and polarization angle (or scan strategy) systematics 
in the presence of beam asymmetry. Then, in order to mitigate the mean-field biases from these 
systematics, we construct bias-hardened estimator analogous to those of our previous work 
\citep{Namikawa:2012pe}. 

\subsection{Non-lensing sources in the off-diagonal covariance} 

\subsubsection{Masking} 

Let us first consider the modification due to a window function, $M(\hatn)$, which is defined to be  
zero for an unmasked region and otherwise unity: 
\al{
	\hT(\hatn) &= [1-M(\hatn)] \tT (\hatn) 
	\,, \\  
	\widehat{P}^\pm(\hatn) &= [1-M(\hatn)] \widetilde{P}^\pm(\hatn) 
	\,. 
}
Such masking mixes E and B modes, leading to mode-coupling in temperature and polarization as 
\al{
	\hT_{\bl} &= \tT_{\bl} - \intl{\bL} M_{\bL}\tT_{\bL'} 
	\,, \\ 
	\hE_{\bl} &= \tE_{\bl} - \intl{\bL} M_{\bL} 
			\left\{\tE_{\bL'} \cos\varphi_{\bL',\bL}-\tB_{\bL'}\sin\varphi_{\bL',\bL}\right\}
	\,, \label{eq:mask E correction} \\ 
	\hB_{\bl} &= \tB_{\bl} - \intl{\bL} M_{\bL} 
			\left\{\tB_{\bL'} \cos\varphi_{\bL',\bL}+\tE_{\bL'}\sin\varphi_{\bL',\bL}\right\}
	\,. \label{eq:mask B correction}
}
With the above equations, the resultant off-diagonal covariance can be written as 
\al{
	\ave{\hX_{\bL}\hY_{\bL'}} &= M_{\bl} f^{M,(XY)}_{\bl,\bL} + \mC{O}(M^2)  \,, 
}
where $\ave{\cdots}$ denotes the usual ensemble average, and the weight functions are summarized in 
Table \ref{table:weight-mask}. The above equation implies that the resultant lensing estimator has 
the mean-field bias due to the mask field, $M_{\bl}$. 

\subsubsection{Inhomogeneous noise/unresolved point-source} \label{Sec:inhomogeneous}

Let us next consider the modification due to addition of arbitrary sky signals, $n^{T}(\hatn)$, $n^{Q}(\hatn)$ and $n^{U}(\hatn)$, which are uncorrelated between pixels -- this approach can be used to model, e.g., residual point sources and inhomogeneous instrumental noise in temperature and polarization maps. The corresponding $\T$, $E$ and $B$ modes are given by
\al{
	n^{\T}_{\bl} &= \FT{\hatn}{\bl}{f} n^{T}(\hatn) 
	\,, \\ 
	n^E_{\bl} \pm \iu n^B_{\bl} &= \FT{\hatn}{\bl}{f} [n^Q \pm \iu n^U ](\hatn) \E^{\mp 2\iu\varphi_{\bl}} 
	\,. 
}
Assuming that $\ave{n^{X}(\hatn)n^{Y}(\hatn')}=S^{(X)}(\hatn)\delta^{XY}\delta(\hatn-\hatn')$, $S^{\T Q}=S^{\T U}=0$ and $S^{(QQ)}=S^{(UU)}\equiv S^{(P)}$, we have
\al{
	\ave{n^\T_{\bL}n^\T_{\bL'}} 
		&= S^{(\T)}_{\bl} 
	\,, \notag \\ 
	\ave{n^E_{\bL}n^E_{\bL'}} 
		&= S^{(P)}_{\bl}\cos 2\varphi_{\bL,\bL'}
	\,, \notag \\
	\ave{n^E_{\bL}n^B_{\bL'}} 
		&= S^{(P)}_{\bl}\sin 2\varphi_{\bL,\bL'}
	\,, \notag \\
	\ave{n^B_{\bL}n^B_{\bL'}} 
		&= S^{(P)}_{\bl}\cos 2\varphi_{\bL,\bL'}
	\,, \notag
}
where we use $\bL+\bL'=\bl$ and define
\al{ 
    S^{(X)}_{\bl} = \FT{\hatn}{\bl}{f} S^{(X)}(\hatn) 
    \,.
}
The off-diagonal covariance then has additional terms 
\al{
	\ave{\hX_{\bL}\hY_{\bL'}} = \ave{n^{X}_{\bL}n^{Y}_{\bL'}} = f_{\bl,\bL}^{S,(XY)}S^{(X)}_{\bl}
	\,. 
}
where the weight function is $f_{\bl,\bL}^{S,(\T\T)}=1$, 
$f_{\bl,\bL}^{S,(EE)}=f_{\bl,\bL}^{S,(BB)}=\cos 2\varphi_{\bL,\bL'}$, and $f_{\bl,\bL}^{S,(EB)}=\sin 2\varphi_{\bL,\bL'}$. 

\subsubsection{Polarization angle with beam asymmetry} 

Instrumental effects such as beam asymmetry and errors in the detector polarization angles are also 
a potential concern for lensing reconstruction. Here we consider the effect of a spatial variation in 
the polarization angle in the presence of the ellipticity in beam shape. 

Denoting CMB temperature and polarization anisotropies as $\Xi^{(0)}=\T$ and 
$\Xi^{(\pm 2)}=Q\pm\iu U=P^{\pm}$, we assume that the beam convolved anisotropies for 
the $i$-th pixel are expressed as follows: 
\al{
	\Xi^{(s)}(\hatn_i) = \Int{2}{\hatn}{}\mC{R}(\hatn_i-\hatn,\polangle(\hatn_i))\Xi^{(s)}(\hatn) 
	\,, \label{Eq:beam-convolved}
}
where $s=0$ (temperature) or $\pm 2$ (polarization), and $\mC{R}$ denotes the beam-response function 
whose shape is independent of the measurements but whose orientation angle, $\polangle(\hatn_i)$ 
(\citealt{Shimon:2007au}), is dependent on both the pixels and measurements. The beam-response 
function, $\mC{R}$, is given by 
\al{
	\mC{R}(\bm{r},\polangle(\hatn_i)) 
		= \Int{2}{\bL}{(2\pi)^2} \E^{\iu\bL\cdot\bm{r}} \mC{R}_{\bL}(\hatn_i)
	\,, 
}
where the Fourier counterpart of the beam-response function is expanded as (\citealt{Shimon:2007au}) 
\al{
	\mC{R}_{\bL}(\hatn_i) 
		= \sum_{n=-\infty}^\infty b_{L,n} \E^{-\iu n\polangle(\hatn_i)}\E^{\iu n \varphi_{\bL}} 
	\,. 
}
Here, by denoting the Bessel function as, $J_n$, the coefficients are given by 
\al{
	b_{L,n} = \iu^n \Int{}{r}{} r J_{n}(Lr) 
		\Int{}{\varphi_r}{2\pi} \mC{R}(\bm{r},0) \E^{-\iu n\varphi_r}
	\,. 
}
Note that $b_{L,-n}=(-1)^nb_{L,n}$, 
and if the shape of beam function, $R(\bm{r},0)$, does not depend on 
the angle, $\varphi_r$, e.g., a circular Gaussian beam, the coefficients are non-zero only when $n=0$. 
The beam-convolved anisotropies are then rewritten as 
\al{
	\Xi^{(s)}(\hatn_i) &= \iu^s \sum_{n=-\infty}^\infty \E^{-\iu n\polangle(\hatn_i)} 
	\notag \\ 
		&\qquad \times \Int{2}{\bL}{(2\pi)^2}b_{L,n}\Xi^{(s)}_{\bL}
			\E^{\iu\bL\cdot\hatn_i}\E^{\iu (n+s) \varphi_{\bL}} 
	\,, 
}
where $\Xi^{(0)}=\T$ and $\Xi^{(\pm 2)}=E\pm\iu B$. 

For a two-beam experiment, as shown in \citet{Shimon:2007au}, the measured temperature and 
polarization anisotropies are distorted by the polarization angle and difference of beam shapes 
between the first and second detector. 
If the anisotropies are measured several times at each pixel, the optimal estimators for temperature 
and polarization anisotropies are given by \citep{Shimon:2007au} 
\al{
	\hXi^{(0)} &= \ave{\Xi^{(0)}_+}\rom{pix} 
		+ \frac{1}{2}\ave{\Xi^{(-2)}_-\E^{2\iu\hpolangle_t}\E^{-2\iu\delta_t}}\rom{pix} 
	\notag \\ 
		&\qquad + \frac{1}{2}\ave{\Xi^{(+2)}_-\E^{-2\iu\hpolangle_t}\E^{2\iu\delta_t}}\rom{pix} 
	\,, \label{Eq:2beam:T} \\ 
	\hXi^{(\pm 2)} &= \ave{\Xi^{(\pm 2)}_+\E^{\pm 2\iu\delta_t}\hPsi^{\pm}_t}\rom{pix} 
		+ \ave{\Xi^{(\mp 2)}_+\E^{\mp 2\iu\delta_t}\E^{\pm 4\iu\hpolangle_t} \hPsi^{\pm}_t}\rom{pix} 
	\notag \\ 
		&\qquad + 2\ave{\Xi^{(0)}_-\E^{\pm 2\iu\hpolangle_t}\hPsi^{\pm}_t}\rom{pix} 
	\,, \label{Eq:2beam:P}
}
where the arguments, $\hatn_i$, are dropped. The bracket, $\ave{\cdots}\rom{pix}$, denotes the average 
over all measurements in each pixel, and $\hpolangle_t$ is the estimated polarization angle for $t$-th 
measurement at each pixel, which has small polarization angle error 
$\delta_t\equiv \hpolangle_t-\polangle_t$. We also define 
\al{
	\hPsi^{\pm 1}_t = \frac{1-\E^{\mp 4\iu\hpolangle_t}\ave{\E^{\pm 4\iu\hpolangle_t}}\rom{pix}}
		{1-\ave{\E^{\mp 4\iu\hpolangle_t}}\rom{pix}\ave{\E^{\pm 4\iu\hpolangle_t}}\rom{pix}} 
	\,. 
} 
The subscripts, $+$ and $-$, in $\Xi^{s}$ are the total and difference of anisotropies 
obtained from the two detectors: 
\al{
	\Xi_+^{(s)} = \frac{\Xi_1^{(s)}\pm \Xi_2^{(s)}}{2}  \,. 
}
Note that, if the temperature and polarization anisotropies are measured only one time for each pixel, 
the quantities, $\hPsi^{\pm 1}_t$, in Eqs.~\eqref{Eq:2beam:T} and \eqref{Eq:2beam:P} should be 
replaced with unity. 

With Eqs.~\eqref{Eq:2beam:T} and \eqref{Eq:2beam:P}, taking into account the polarization 
angle involved in $\Xi^{(s)}$, the measured temperature and polarization in Fourier space are given by 
\al{
	\hX_{\bl} &= \sum_{n=-\infty}^\infty \sum_{p=0,\pm 1}\sum_{Y=\T,E,B} \intl{\bL}
	\notag \\ 
		&\qquad \times \Beam^{(XY)}_{L,(n,p)} \tY_{\bL} \pols^{(n,p)}_{\bL'}
			\E^{\iu n\varphi_{\bL,\bL'}}\E^{2\iu p\varphi_{\bL,\bl}}
	\,, \label{eq:beam-conv}
}
where $X=\T$, $E$ or $B$, the quantities, $\pols^{(n,p)}_{\bL}$ ($p=0,\pm 1$), are defined as 
the Fourier transform of the following quantities: 
\al{
	\pols^{(n,p)}(\hatn_i) &= \ave{\E^{-\iu n \hpolangle_t(\hatn_i)}
		\E^{\iu (n+2p)\delta_t(\hatn_i)}\hPsi^{p}_t(\hatn_i)}\rom{pix} \E^{\iu n\varphi_{\bL}} 
	\,, \label{Eq:pol-fourier}
}
with $\hPsi^0_t=1$. Note that $\pols^{(n,p)}_{\bL}$ is the spin-$(-n)$ transform of spin-$(-n)$ 
quantity, $\ave{\E^{-\iu n \hpolangle_t(\hatn_i)}\hPsi^{p}_t(\hatn_i)}\rom{pix}$, in the limit of 
$\delta_i=0$ for all measurements. The coefficients, $\Beam^{(ZZ')}_{L,(n,p)}$, are given by 
\al{
	&\Beam^{(\T\T)}_{L,(n,0)} = b^+_{L,n}  \notag \\ 
	&\Beam^{(\T E)}_{L,(n,0)} = -\frac{b^-_{L,n+2} + b^-_{L,n-2}}{2}  \notag \\ 
	&\Beam^{(\T B)}_{L,(n,0)} = \iu \frac{b^-_{L,n+2} - b^-_{L,n-2}}{2}  \notag \\ 
	&\Beam^{(E\T)}_{L,(n,\pm 1)} = -b^-_{L,n\pm 2}  \notag \\ 
	&\Beam^{(EE)}_{L,(n,\pm 1)} = \frac{b^+_{L,n} + b^+_{L,n\pm 4}}{2}  \notag \\ 
	&\Beam^{(EB)}_{L,(n,\pm 1)} = \pm \iu \frac{b^+_{L,n} - b^+_{L,n\pm 4}}{2}  \notag \\
	&\Beam^{(BY)}_{L,(n,\pm 1)} = \mp\iu \Beam^{(EY)}_{L,(n,\pm 1)}  \,, 
}
where $b^\pm_{L,n}$ is the total and difference of beam transfer functions for two detectors: 
\al{
	b^\pm_{L,n} = \frac{b^{(1)}_{L,n}\pm b^{(2)}_{L,n}}{2} 
	\,. 
}
Note that, in full sky and only for temperature case, Eq.~\eqref{eq:beam-conv} is consistent 
with \citet{Hanson:2010gu}. 

In Fourier space, we break $\pols^{(n,p)}_{\bl}$ into constant and fluctuation pieces
\al{
	\pols^{(n,p)}_{\bl} = C^{(n,p)}\delta_{\bl=\bm{0}} + (\pols^{(n,p)}_{\bl})\rom{ani}  \,, 
}
with the assumption that $(\pols^{(n,p)}_{\bl})\rom{ani}$ are small. 
Then, Eq.~\eqref{eq:beam-conv} is rewritten as 
\al{
	\hX_{\bl} &= \sum_{n=-\infty}^{\infty}\sum_{p=0,\pm 1}\sum_{Y=\T,E,B} 
		\bigg\{ \Beam^{(XY)}_{,(n,p)} \tY_{\bl}\E^{\iu n\varphi_{\bl}} C^{(n,p)}
	\notag \\ 
		+& \intl{\bL} \Beam^{(XY)}_{L,(n,p)} \tY_{\bL} (\pols^{(n,p)}_{\bL'})\rom{ani}
			\E^{\iu n\varphi_{\bL,\bL'}}\E^{2 \iu p\varphi_{\bL,\bl}}\bigg\}
	\,, \label{eq:beam-conv-decomp}
}
For realistic cases, $b^+_{L,n}/b^+_{L,0}\ll 1$ for $n\not=0$, $b^-_{L,n}\ll b^+_{L,n}$ and 
$\delta_i\ll 1$. Under these approximations, the dominant term in the first term of 
Eq.~\eqref{eq:beam-conv-decomp} becomes $b^+_{,0}X_{\bl}$. Assuming that $b^+_{,0}=1$, the 
convolution in Eq.~\eqref{eq:beam-conv-decomp} leads to an off-diagonal covariance given by 
\al{
	\ave{\hX_{\bL}\hY_{\bL'}} &= \sum_{n\not=0}\sum_{p=0,\pm 1}
		f^{\pols^{(n,p)},(XY)}_{\bl,\bL}(\pols^{(n,p)}_{\bl})\rom{ani} 
	\notag \\ 
		&\qquad\qquad + \mC{O}[(\pols^{(n,p)})^2\rom{ani}] 
	\,, 
}
where the weight functions are summarized in Table \ref{table:weight-beam}. 
Since $\pols^{(n,p)}$ is a spin-$n$ quantities, it is useful to define a rotational-invariant quantities: 
\al{
	\pols^{(n,e)}_{\bl} &= \pols^{(-n,0)}_{\bl} + (-1)^n\pols^{(n,0)}_{\bl}  \,, \\
	\pols^{(n,b)}_{\bl} &= -\iu [\pols^{(-n,0)}_{\bl} - (-1)^n\pols^{(n,0)}_{\bl}]  \,, \\
	\pols^{(n,\varepsilon)}_{\bl} &= \pols^{(-n,+1)}_{\bl} + (-1)^n\pols^{(n,-1)}_{\bl}  \,, \\
	\pols^{(n,\beta)}_{\bl} &= -\iu [\pols^{(-n,+1)}_{\bl} - (-1)^n\pols^{(n,-1)}_{\bl}]  \,, 
}
where the above quantities satisfy $(\pols^{(n,\varepsilon)}_{\bl})^*=\pols^{(n,\varepsilon)}_{-\bl}$ 
and $(\pols^{(n,\beta)}_{\bl})^*=\pols^{(n,\beta)}_{-\bl}$. 
The corresponding weight function is then given as
\al{
	f^{(n,e)}_{\bl\bL} &= f^{(-n,0)}_{\bl\bL} + (-1)^nf^{(n,0)}_{\bl\bL}  \,, \\
	f^{(n,b)}_{\bl\bL} &= -\iu [f^{(-n,0)}_{\bl\bL} - (-1)^nf^{(n,0)}_{\bl\bL}]  \,, \\
	f^{(n,\varepsilon)}_{\bl\bL} &= f^{(-n,+1)}_{\bl\bL} + (-1)^nf^{(n,-1)}_{\bl\bL}  \,, \\
	f^{(n,\beta)}_{\bl\bL} &= -\iu [f^{(-n,+1)}_{\bl\bL} - (-1)^nf^{(n,-1)}_{\bl\bL}]  \,, 
}
%
For $b^+_{,0}\not=1$, we can utilize $f^{\pols^{(n,p)},(XY)}_{\bl,\bL}/(b^+_{L,0}b^+_{L',0})$ 
for the weight function. Note that the above derivations cover simpler cases; 
if the beam of two detectors are the same, $b^-=0$, with Gaussian shape, 
$b_{L,n}\propto \delta_{n,0}$, and $\hpolangle_t$ is the same for all measurements, we obtain 
$\hXi^{(\pm 2)}(\hatn) = \Xi^{(\pm 2)}(\hatn)\E^{\pm 2\iu \delta(\hatn)}$. 
We also note that, for only temperature case, the results are consistent with our previous work. 

\subsection{Mean-field biases} 

All of the above contaminations lead to the mean-field bias for lensing estimator, 
$\estx_{\bl}^{(XY)}$. Omitting the subscript $(XY)$, the mean-field biases are given by 
\al{
	\ave{\estx_{\bl}} &= A_\l^{xx} \intl{\bL} g^x_{\bl,\bL} \ave{\bX_{\bL}\bY_{\bL'}}  \notag \\ 
		&= A_\l^{xx} \intl{\bL} \sum_{y}g^x_{\bl,\bL}f^y_{\bl,\bL} y_{\bl} \notag \\ 
		&= \sum_y R^{xy}_\l y_{\bl} 
	\,, 
}
where $y=M,S$ or $\pols^{(n,p)}$, and we define the response function $R_\l$ and normalization $A_\l$ 
as: 
\al{
	R_\l^{xy} &= \frac{A_\l^{xx}}{A_\l^{xy}} \,; \quad 
	A_\l^{xy} &= \left\{\intl{\bL} g^x_{\bl,\bL} f^y_{\bl,\bL'}\right\}^{-1} 
	\,. \label{Eq:response}
}

\subsection{Bias-hardened estimator} 

Estimators which are bias-hardened against the effects above many be constructed analogously to 
the temperature case, i.e., we first construct a naive estimator for a given effect $y$ as
\al{
	\esty_{\bl} = A_\l^{yy} \intl{\bL} g^y_{\bl,\bL} \bX_{\bL}\bY_{\bL'}  \,, 
}
where $g^y_{\bl,\bL}$ and $A_\l^{yy}$ are defined as Eq.~\eqref{Eq:def-gXY} and \eqref{Eq:def-AXY}, 
respectively, but using the weight function, $f^{y,(XY)}_{\bl,\bL}$, instead of the lensing weight 
function, $f^{x,(XY)}_{\bl,\bL}$ $(x=\grad,\curl)$. This estimator for $y_\l$ is in turn biased by 
lensing. We may then obtain a bias-hardened estimator as
\al{
	\estx^{(\rm BHE)}_{\bl} \equiv \sum_y \{\bR{R}_\l^{-1}\}^{x,y}\esty_{\bl}  \,. \label{BR-est}
}

\section{Demonstration of bias-hardened estimator for mean-field bias} \label{sec.4}

\begin{figure*}
\bc
\includegraphics[width=80mm,clip]{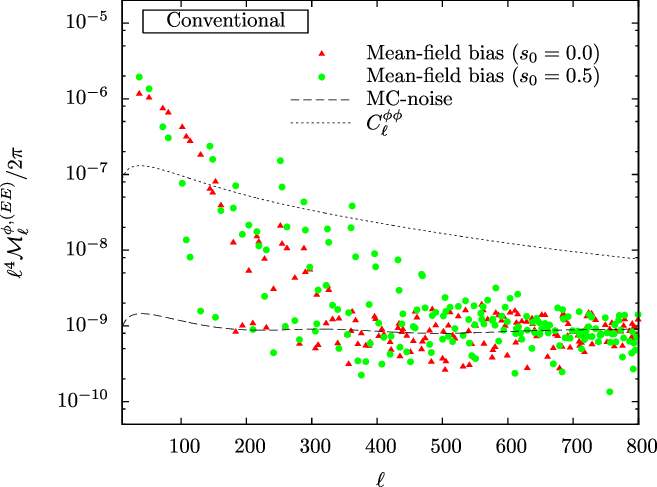} 
\includegraphics[width=80mm,clip]{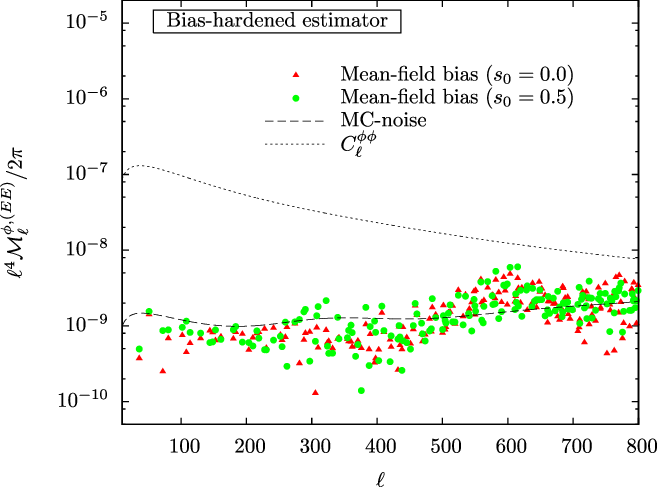}
\caption{
{\it Left}: Mean-field power spectrum for the gradient-mode $EE$ estimator, $\mC{M}^{\grad,(EE)}_\l$, 
estimated with $100$ lensed simulated maps. The left panel shows results with the conventional 
estimator, varying the apodization parameter, $s_0$, as $0.0$ and $0.5$. 
The Monte-Carlo noise floors (dashed line) are shown comparing with the mean-field power spectrum. 
The theoretical lensing power spectrum is also shown as a solid gray line. 
Note that the reconstruction is performed on $(5\deg)^2$, with $E$-mode multipoles ranges of, 
$2\leq\leq 3000$. 
{\it Right}: Same as left panel but with the bias-hardened estimator. 
}
\label{Fig:EE:grad}
\ec
\end{figure*}

\begin{figure*}
\bc
\includegraphics[width=80mm,clip]{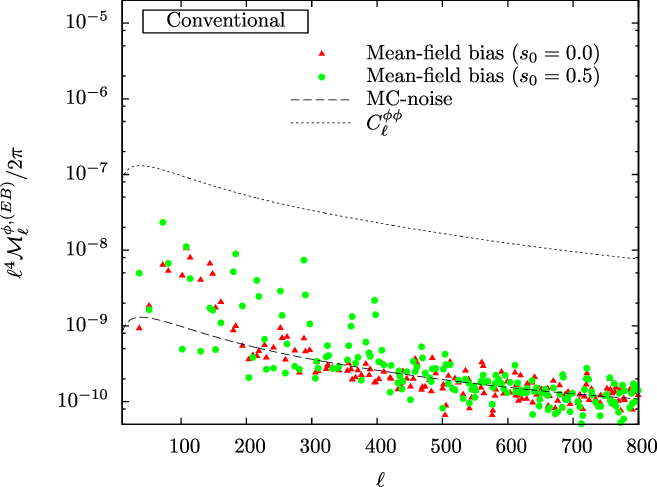} 
\includegraphics[width=80mm,clip]{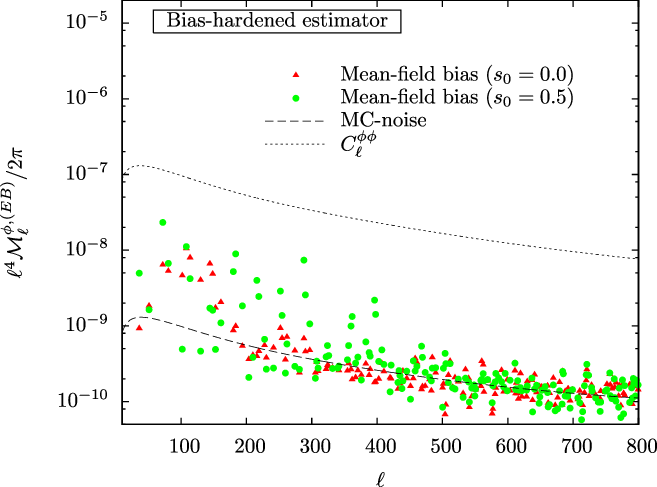}
\caption{
Same as Fig. \ref{Fig:EE:grad}, but for the $EB$ case. 
}
\label{Fig:EB:grad}
\ec
\end{figure*}

\begin{figure*}
\bc
\includegraphics[width=80mm,clip]{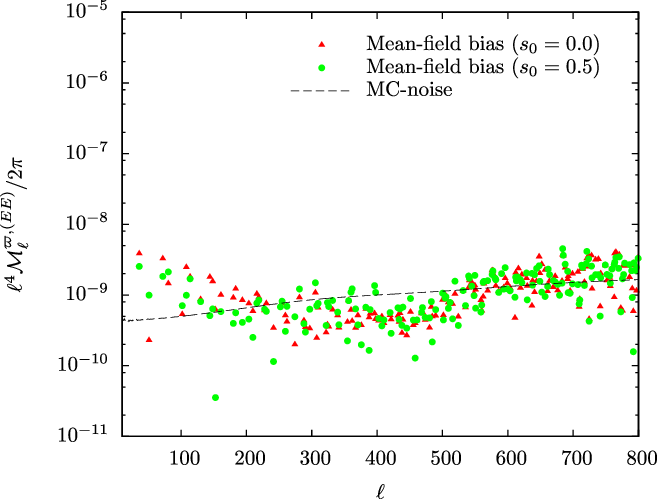} 
\includegraphics[width=80mm,clip]{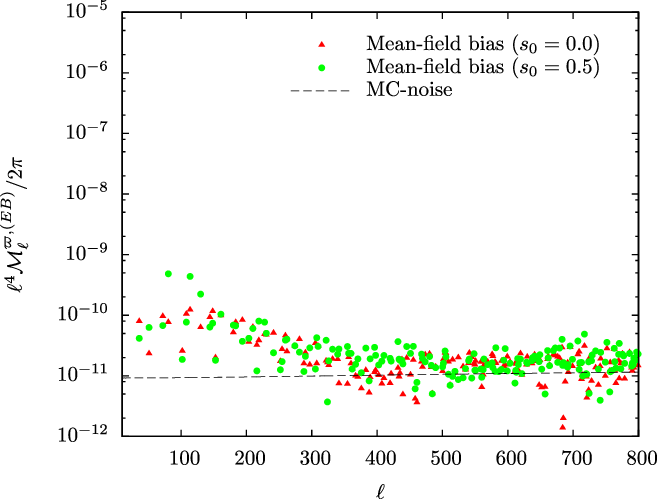} 
\caption{
Mean-field power spectrum for curl modes with $EE$ (left) and $EB$ (right) estimators, respectively. 
The Monte-Carlo noise floors are shown with dashed lines. 
}
\label{Fig:curl}
\ec
\end{figure*}

\begin{figure*}
\bc
\includegraphics[width=80mm,clip]{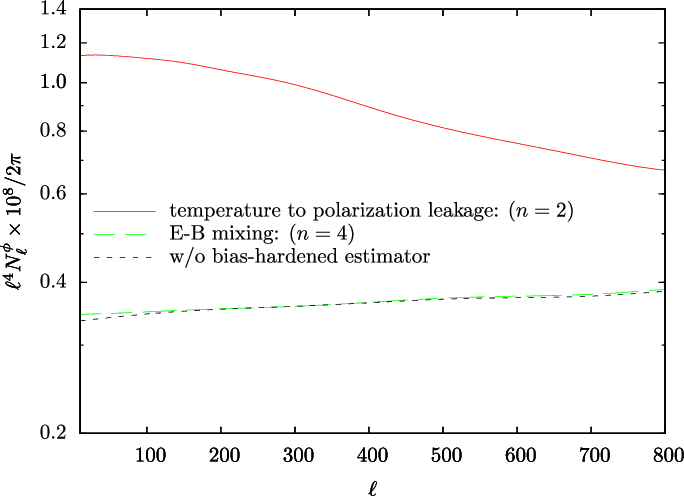} 
\includegraphics[width=80mm,clip]{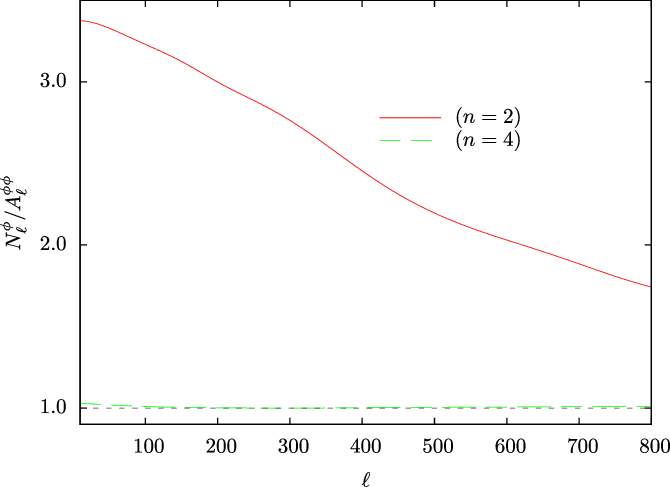} 
\caption{
{\it Left}: Comparison of the noise level of lensing reconstruction between the case with and without 
the bias-hardened estimator for $EB$ case. The noise level of the bias-hardened estimator, 
$N_\l^\grad$, is computed from Eq.~\eqref{Eq:noise-level}. The labels, $n=2$ and $n=4$, denote the 
temperature to polarization leakage, and $E$-$B$ mixing by the rotation of the coordinate system, 
respectively (see text for the definitions and details). 
{\it Right}: The ratio of the noise level described in the left panel compared to the case w/o 
the bias-hardened estimator, $N_\l^\grad/A_\l^{\grad\grad}$. 
}
\label{Fig:Polangle}
\ec
\end{figure*}

In this section, we discuss whether the bias-hardened estimator for lensing fields can be used as a 
cross-check for conventional estimator. For this purpose, we first compute the case where the mean 
field is generated only from the effect of masking. One concern here is the validity of linear-order 
approximation. That is, to derive bias-hardened estimator, we have ignored any higher-order terms of 
$M_{\bL}$ (and, of course, for other non-lensing fields). 

\subsection{Simulated Maps and Analysis} 

We use simulated polarization maps produced using methods similar to our previous work 
\citep{Namikawa:2012pe}. For lensing reconstruction, we use $100$ realizations of lensed Stokes $Q$ 
and $U$ maps, simulated on a $5\times 5$ deg$^2$ patch. The details of the method used to generate 
these lensed maps are described in Appendix~\ref{app:sim}. To simulate the masking of point sources we 
create masks by cutting $200$ regions of randomly located $10'\times 10'$ squares. Note that the area 
covered by the point-source masks is $\sim 5$ \% of total area, and the percentage roughly corresponds 
to that used in the SPT polarization analysis \citep{Hanson:2013daa}. To consider experiments with 
high-angular resolution such as SPTpol, PolarBear and ACTPol, as well as to avoid contamination by the 
Sunyaev-Zel'dovich effect \citep{Zeldovich:1969} and unresolved point sources, we assume a delta 
function instrumental beam. The $E$ and $B$ modes multipoles are used at $2\leq\leq3000$. We 
assume homogeneous map noise, with a level of 0.01 $\mu$K-arcmin. Note that, even in the presence of 
an inhomogeneous noise, by combining the bias-hardened estimator described in 
Sec.~\ref{Sec:inhomogeneous}, the mean field due to inhomogeneous noise would be reduced as already 
applied to the lensing reconstruction with Planck temperature map \citep{Ade:2013tyw}, and the 
qualitative result would be similar to that obtained in this paper. 

\subsection{Filtering}

For the conventional estimator approach, we experiment with the following filtering techniques to 
suppress the mask-mean field: apodization of the survey boundary and $C^{-1}$ filtering for the 
point-source holes. 

\subsubsection{Apodization window function} 

In our analysis, we use the following analytic apodization function whose value and derivatives are 
zero at the boundaries of the survey region: 
\al{
	W(x,y;s_0) &= w(x;s_0)w(y;s_0)M(x,y)  \,, 
}
where $w(s;s_0)$ is a sine apodization function given by
\al{
	w(s;s_0) &= \begin{cases} 1 & |s|<as_0 \\ 
		\dfrac{1-|s|/a}{1-s_0} - \dfrac{1}{2\pi}\sin\left(2\pi\dfrac{1-|s|/a}{1-s_0}\right) 
			& as_0\leq |s|<a \\ 0 & a\leq |s| \end{cases}
	\,, \label{apo}
}
and $M(x,y)$ represents the point-source mask, i.e., $0$ at the presence of (resolved) point sources, 
and otherwise $1$. The parameter, $s_0$, indicates the width of the region where the apodization is applied. 

\subsubsection{$C^{-1}$ filtering} 

The minimum-variance filtering which emerges from likelihood-based derivations of lensing estimators 
is known as $C^{-1}$ filtering. The inverse-variance filtered Fourier modes, 
$\bar{\bm{X}}_{\bl}=(E_{\bl},B_{\bl})$, are obtained by solving
\al{
	\left[1+\bR{C}^{1/2}\bR{N}^{-1}\bR{C}^{1/2}\right] (\bR{C}^{1/2}\bar{\bm{X}}) 
		= \bR{C}^{1/2}\bR{N}^{-1} \hat{X}
	\,, \label{Cinv}
}
where $\bar{\bm{X}}$ is a vector whose components are $\bar{X}_{\bl}$, $\bR{C}$ is the covariance 
of the CMB anisotropies with 
\al{
	\{\bm{C}\}_{\bl_i,\bl_j} &= \delta_{\bl_i-\bl_j} \Mat{ C_{\l_i}^{EE} & 0 \\ 0 & C_{\l_i}^{BB} }
	\,, 
}
and $\bm{N}=\ave{\bm{n}^{\dagger}\bm{n}}$ is the covariance matrix for the instrumental noise. 
The noise covariance matrix in Fourier space is obtained from that in real space as 
\al{
	\bm{N}^{-1} = \bm{Y}^{\dagger}\bm{\ol{N}}^{-1}\bm{Y}  \,, 
}
where the pointing matrix, $\bm{Y}$, is defined by 
\al{
	\{\bm{Y}\}_{\hatn_i,\bl_j} = \exp(\iu\hatn_i\cdot\bl_j)\exp(-2\varphi_{\bl_j})\Mat{1&\iu\\1&-\iu}
	\,. 
}
Note that the matrix in the above equation describes the transformation of $(E_{\bl},B_{\bl})^t$ to 
$(E_{\bl}+\iu B_{\bl},E_{\bl}-\iu B_{\bl})^t$. The mask is incorporated by setting the noise level of 
masked pixels to infinity, and therefore the inverse of the noise covariance in real space 
$\bm{\ol{N}}^{-1}$ to zero for masked pixels. The inversion of the matrix on the left-hand side of 
Eq.~(\ref{Cinv}) can be numerically costly, but may be evaluated using conjugate descent with careful 
preconditioning \citep{Smith07}. Since the mask mean-filed due to the survey boundary remains after 
applying the $C^{-1}$ filter, we additionally apply an apodizing function given by Eq.~(\ref{apo}). 

\subsection{Mean-field bias due to masking} 

We now turn to discuss the mean-field bias generated by masking. 

Given $N$ realizations of estimator, $\estx_{\bl}^{i,(XY)} \ (i=1,2,\dots,N)$, we define the 
mean-field power spectrum as 
\al{
	\mC{M}^{x,(XY)}_\l &= \frac{1}{W_4}\Int{}{\varphi_{\bl}}{2\pi}
		\bigg|\frac{1}{N}\sum_{i=1}^N\estx_{\bl}^{i,(XY)}\bigg|^2 
	\,, 
}
where the quantity $W_4$ is the normalization correction for effect of window function as 
\al{
	W_4 = \Int{2}{\hatn}{} [W(\hatn)]^4  \,. 
}
With $N$-realizations of CMB maps, the mean-field becomes \citep{BenoitLevy:2013bc}
\al{
	\mC{M}^{x,(XY)}_\l &\simeq \frac{1}{W_4}\Int{}{\varphi_{\bl}}{2\pi}|\ave{\estx_{\bl}^{(XY)}}|^2 
		+ \frac{A_\l^{xx,(XY)}+C_\l^{xx}}{N} 
	\,. \label{Eq:MC-noise}
}

The resulting mask mean-field for $\estg$ are shown in Figs.~\ref{Fig:EE:grad} and \ref{Fig:EB:grad} 
for quadratic combination of $EE$ and $EB$ cases, respectively. We compare results between the case 
with and without the bias-hardened estimator for mitigating mask mean-field. We also vary the 
apodization parameter, $s_0$, introduced in Eq.~(\ref{apo}). The Monte-Carlo noise floor is the second 
term of Eq.~\eqref{Eq:MC-noise}. For $EE$, the mask mean-field is large on large angular scales, and 
exceeds the expected lensing power. We can clearly see that the bias-hardened estimator works well to 
mitigate the mask mean-field. On the other hand, the $EB$ estimator has small contributions to the 
mask mean-field. The reason is as follow. If we use the filtering methods, the quantity $M_\l$ has 
value only around $\sim 0$, and thus the mask-mean field, which is expressed as $R^{xM}_\l M_\l$, is 
significant only for $\sim 0$. For $EB$ estimator, however, the response function, $R^{xM}_\l$, given 
in Eq.~(\ref{Eq:response}), has the sine function involved in the weight function, 
$f_{\bl,\bL}^{x,(EB)}$, and goes to zero as $\to 0$, which is contrary to the case of $EE$ estimator 
in which the weight function includes the cosine function instead of sine function. Note that the 
origin of the sine function in the weight function is the parity of $EB$ cross-correlation, i.e., odd 
parity symmetry, as the weight function is given by Eq.~\eqref{Eq:weight}. We have also checked that, 
since the behavior of the response function for $\T$B is similar to that of $EB$, the mask mean-field 
is negligible for $\T$B estimator. As expected, the mask mean-field on the curl mode shown in 
Fig.~\ref{Fig:curl} would also be negligible for both $EE$ and $EB$ estimator. 

The residual mean-field bias also affects on the power spectrum estimation through 
\al{
	\Int{}{\varphi_{\bl}}{2\pi}|\estx_{\bl}|^2 = \mC{M}^x_\l + A_\l^{xx} + C_{}^{xx}  \,. 
}
The figure shows that, for $EE$, the conventional estimator generates a significant mask-mean field 
which exceeds the lensing power spectrum;  the lensing power spectrum estimation could therefore 
suffer from uncertainties in the mean-field correction. On the other hand, the bias-hardened estimator 
significantly suppresses the mask-mean field which is below the lensing power spectrum. 

\subsection{Mean-field bias due to polarization angle} 

The mask-mean field for the $EB$ estimator is negligible even for the case with a simple apodization 
function, however, the mean-field biases can be generated from other sources such as polarization 
angle errors. Here, to see the potential of bias-hardened estimators to mitigate polarization angle 
systematics, we compute a rough estimation of the response function and degradation of noise level for 
$EB$ estimator. 

For a two-beam experiment, with $b^-\not=0$ and $b_{L,n}\propto \delta_{n,0}$, 
as an example, we consider the following non-zero weight functions for $EB$ estimator: 
\al{
	f^{\pols^{(\pm 2,\mp 1)}}_{\bl,\bL} &= \mp\iu\E^{\pm 2\iu\varphi_{\bL',\bl}} b^-_{L,0}\tCTE_L 
	\\ 
	f^{\pols^{(\pm 4,\mp 1)}}_{\bl,\bL} &= \pm\iu\E^{\pm 2\iu(\varphi_{\bL,\bl}+\varphi_{\bL',\bl})} 
		(b_{L,0}\tCBB_L + b_{L',0} \tCEE_{L'}) 
	\,, 
}
where we have omitted the label, $(EB)$, in the weight functions. Note that the cases with 
$(n,p)=(\pm 2,\mp 1)$ and $(\pm 4,\mp 1)$ denote the systematics due to temperature to polarization 
leakage and E-B modes mixing, respectively. In general, as an alternative to $\pols^{(n,p)}_{\bl}$, 
we can use the following quantities:  
\al{
	\pols^{(n,\varepsilon)}_{\bl} &= \pols^{(-n,+1)}_{\bl} + (-1)^n\pols^{(n,-1)}_{\bl}  \,, \\
	\pols^{(n,\beta)}_{\bl} &= -\iu [\pols^{(-n,+1)}_{\bl} - (-1)^n\pols^{(n,-1)}_{\bl}]  \,, 
}
where the above quantities satisfy $(\pols^{(n,\varepsilon)}_{\bl})^*=\pols^{(n,\varepsilon)}_{-\bl}$ 
and $(\pols^{(n,\beta)}_{\bl})^*=\pols^{(n,\beta)}_{-\bl}$. The corresponding weight functions for 
$\pols^{(n,\varepsilon)}_{\bl}$ and $\pols^{(n,\beta)}_{\bl}$ are 
\al{
	f^{\pols^{(2,\varepsilon)}}_{\bl,\bL} &= b^-_{L,0}\sin(2\varphi_{\bL',\bl})\tCTE_L 
	\,, \notag \\ 
	f^{\pols^{(2,\beta)}}_{\bl,\bL} &= -b^-_{L,0}\cos(2\varphi_{\bL',\bl})\tCTE_L 
	\,, \notag \\ 
	f^{\pols^{(4,\varepsilon)}}_{\bl,\bL} 
		&= -\sin(2\varphi_{\bL,\bl}+2\varphi_{\bL',\bl})(b_{L,0}\tCBB_L + b_{L',0} \tCEE_{L'}) 
	\,, \notag \\ 
	f^{\pols^{(4,\beta)}}_{\bl,\bL} 
		&= \cos(2\varphi_{\bL,\bl}+2\varphi_{\bL',\bl})(b_{L,0}\tCBB_L + b_{L',0} \tCEE_{L'}) 
	\,. \label{Eq:n=2,4}
}
In our calculation, for simplicity, we assume $b^-_{L,0}\equiv \epsilon b_{L,0}$ and a top-hat 
function for $b_{L,0}$, i.e., $b_{L,0}=1$ for $\leq 3000$ and $0$ otherwise, and also ignore $\tCBB_L$ 
in the above equation. Since the weight functions of $\pols^{(n,\beta)}_{\bl}$ is obtained only by 
replacing sine function in $f_{\bl,\bL}^{(n,\varepsilon)}$ with cosine function, we expect that the 
amplitude of the noise level for $\pols^{(n,\beta)}_{\bl}$ would be not so different from that for 
$\pols^{(n,\varepsilon)}_{\bl}$, and we only focus on the case to mitigate 
$\pols^{(n,\varepsilon)}_{\bl}$ in the following calculations. 

In the left panel of Fig.~\ref{Fig:Polangle}, we show the noise level for the bias-hardened estimator 
incorporating polarization angle systematics of $\pols^{(n,\varepsilon)}_{\bl}$. The noise level for 
the conventional approach corresponds to the normalization, $A^{\grad\grad}_\l$. On the other hand, 
with $n=2$ and $4$, the noise level for the bias-hardened estimator is given by 
\al{
	N_\l^\grad = \frac{A^{\grad\grad}_\l}
		{1-R^{\grad \pols^{(n,\varepsilon)}}_\l R^{\pols^{(n,\varepsilon)} \grad}_\l} 
	\,, \label{Eq:noise-level}
}
where the response functions, $R^{\grad \pols^{(n,\varepsilon)}}$ and 
$R^{\pols^{(n,\varepsilon)}\grad}$, defined in \eq{Eq:response}, are computed with the weight 
functions given in \eq{Eq:n=2,4}. Note that the noise level does not depend on $\epsilon$. We 
find that the noise level is not necessarily much larger using a bias-hardened estimator compared to 
the conventional approach. In the right panel, we also show the ratio of the case with the 
bias-hardened estimator to that with conventional estimator. We find that the degradation of noise 
level is only up to $\alt 1\%$ for $n=4$, and by a factor of $\sim 3$ for $n=2$. Our results imply 
that the bias-hardened estimator is enough to utilize for a cross check of usual method for 
polarization angle systematics. 

\section{Summary} \label{sec.5} 

We have discussed methods for mitigating the mean-field bias in the case of lensing reconstruction 
with CMB polarization. We first derived the mean-field bias generated from masking, inhomogeneous 
noise (and/or unresolved point sources), and polarization angle systematics associated with the 
asymmetric beam shape, in analogy to the temperature-only case. Then we performed numerical tests to 
see how significantly the mean-field bias from masking is mitigated with the bias-hardened estimator. 
We found that, for $EE$ estimator, it particularly useful for the reduction of the large-scale 
component of the mean-field. On the other hand, for the $EB$ estimator, we found that the amplitude of 
mask mean-field is negligible compared to the lensing signal. The bias-hardened $EB$ estimator, 
however, is useful for other potential sources of mean field, such as polarization angle systematics, 
and we showed that the increase of noise level is only up to $1\%$ for $n=4$ ($E$-$B$ mixing), and by 
a factor of $\sim 3$ for $n=2$ (temperature to polarization leakage), compared to the conventional 
approach.

\section*{Acknowledgments}

We greatly appreciate Duncan Hanson for valuable comments and helpful discussions. We are also 
grateful to Takashi Hamana and Takahiro Nishimichi for kindly providing the ray-tracing simulation 
code and the 2LPT code, and thank Ryo Nagata for useful comments. This work was supported in part by 
Grant-in-Aid for Scientific Research on Priority Areas No. 467 ``Probing the Dark Energy through an 
Extremely Wide and Deep Survey with Subaru Telescope'', by the MEXT Grant-in-Aid for Scientific 
Research on Innovative Areas  (No. 22111501), by JSPS Grant-in-Aid for Research Activity Start-up 
(No. 80708511), and by JSPS Grant-in-Aid for Scientific Research (B) (No. 25287062) 
``Probing the origin of primordial mini-halos via gravitational lensing phenomena''. 
Numerical computations were carried out on SR16000 at YITP in Kyoto University and Cray XT4 at Center 
for Computational Astrophysics, CfCA, of National Astronomical Observatory of Japan.

\appendix

\section{Bias-hardened estimator for lensing power spectrum} \label{app:BHE} 

Here we present an optimal estimator for the lensing angular power spectrum, $\widehat{C}^{xx}_\l$, 
motivated by the maximum likelihood estimator for lensing trispectrum, as proposed in our previous 
work \citep{Namikawa:2012pe} where we considered the temperature anisotropies alone. 

\subsection{Formalism} 

\subsubsection{Likelihood for lensed CMB anisotropies} 

Gaussian probability distribution function for temperature and polarization fields, $a=\T$, $E$ or 
$B$, whose covariance matrix are ${\rm\bm{C}}^{a_{\bl},b_{\bl'}}=\ave{a_{\bl}b_{\bl'}}$, is given by 
\al{
	P\rom{g} = \frac{1}{\sqrt{(2\pi)^N\det C}} \exp\left(-\frac{1}{2}\sum_{ab}\sum_{\bl,\bl'}
		a_{\bl}({\rm\bm{C}}^{-1})^{a_{\bl},b_{\bl'}}b_{\bl'}\right) 
	\,, 
}
Since the lensed anisotropies, $\tT,\tE,\tB$, are no longer the Gaussian fields, the perturbative 
expansion of the likelihood for the lensed anisotropies at leading order is given as 
\citep{Amendola:1996,Regan:2010cn}
\al{
	P = \left[1+\sum_{abcd}\sum_{\bl_i}\ave{a_{\bl_1}b_{\bl_2}c_{\bl_3}d_{\bl_4}}\rom{c}
		\PD{}{a_{\bl_1}}\PD{}{b_{\bl_2}}\PD{}{c_{\bl_3}}\PD{}{d_{\bl_4}}\right] P\rom{g} 
	\,. \label{Eq:likelihood}
}
Here, we ignore the three-point correlation because 
this is generated due to the correlation between the integrated Sachs-Wolfe effect and lensing.
The cumulant is given by 
\al{
	\ave{a_{\bl_1}b_{\bl_2}c_{\bl_3}d_{\bl_4}}\rom{c}
	&\simeq f^{ab}_{\bl_{12},\bl_1}f^{cd}_{-\bl_{12},\bl_3}C^{\grad\grad}_{|\bl_{12}|}\delta_{\bl_{12},-\bl_{34}} 
	\notag \\ 
	&\quad + f^{ac}_{\bl_{13},\bl_1}f^{bd}_{-\bl_{13},\bl_2}C^{\grad\grad}_{|\bl_{13}|}\delta_{\bl_{13},-\bl_{24}} 
	\notag \\ 
	&\quad + f^{ad}_{\bl_{14},\bl_1}f^{bc}_{-\bl_{14},\bl_2}C^{\grad\grad}_{|\bl_{14}|}\delta_{\bl_{14},-\bl_{23}}
	\,. \label{Eq:cumulant} 
}
Substituting \eq{Eq:cumulant} into \eq{Eq:likelihood}, we obtain the probability 
distribution function for lensed CMB anisotropies. Note here that, we do not compute higher order 
terms of $C^{\grad\grad}_\l$, since we use an approximation which requires the expression only up to 
the first order of $C^{\grad\grad}_\l$. 

\subsubsection{Derivative of probability distribution function} 

To obtain the maximum-likelihood point, we differentiate $P$ with respect to $C^{xx}_\l$, and obtain 
\al{
	\PD{P}{C^{xx}_\l} &= \sum_{abcd}(\widehat{f}_{\bl}^{ab}\widehat{f}_{-\bl}^{cd} 
		+ \widehat{f}_{\bl}^{ac}\widehat{f}_{-\bl}^{bd} 
		+ \widehat{f}_{\bl}^{ad}\widehat{f}_{-\bl}^{bc})P\rom{g}
	\,, \label{Eq:likelihood}
}
where the operator is defined as 
\al{
	\widehat{f}_{\bl}^{ab} \equiv \sum_{\bl_1}f^{ab}_{\bl,\bl_1}\PD{}{a_{\bl_1}}\PD{}{b_{\bl-\bl_1}} 
	\,. 
}
We find that 
\al{
	\widehat{f}_{\bl}^{ab}P\rom{g} = (\barx^{ab}_{\bl}-\ave{\barx^{ab}_{\bl}})P\rom{g}
	\,, \label{eq:f-operate}
}
where we define the unnormalized estimator as 
\al{
	&\barx^{ab}_{\bl} = \sum_{\bl_1}f^{ab}_{\bl,\bl_1}\ol{a}_{\bl_1}\ol{b}_{\bl-\bl_1} 
	\,, 
}
with the inverse variance filtered multipoles as 
$\ol{a}_{\bl}=\sum_{a',\bl'}({\bR{C}}^{-1})^{a_{\bl}a'_{\bl'}}a'_{\bl'}$. 
Operating $\widehat{f}_{-\bl}^{cd}$ again to \eq{eq:f-operate}, we obtain 
\al{
	\frac{1}{P\rom{g}}\widehat{f}_{-\bl}^{cd}\widehat{f}_{\bl}^{ab}P\rom{g} 
		&= \barx^{ab,(C)}_{\bl}\barx^{cd,(C)}_{-\bl} - n_{\bl}^{ab,cd}
	\,, \label{Eq:fabcd-derive}
}
where the mean-field corrected estimator and its reconstruction noise bias are given by 
\al{
	& \barx^{ab,(C)}_{\bl} \equiv \barx^{ab}_{\bl} - \ave{\barx^{ab}_{\bl}} 
	\\ 
	& n_{\bl}^{ab,cd} \equiv \ave{(\barx^{a^{(1)}b}_{\bl}+\barx^{ab^{(1)}}_{\bl})
		(\barx^{c^{(1)}d}_{-\bl}+\barx^{cd^{(1)}}_{-\bl})}_{(1)}
	\notag \\ 
		&\quad - \frac{1}{2}\ave{(\barx^{a^{(1)}b^{(2)}}_{\bl}+\barx^{a^{(2)}b^{(1)}}_{\bl})
		(\barx^{c^{(1)}d^{(2)}}_{-\bl}+\barx^{c^{(2)}d^{(1)}}_{-\bl})}_{(1),(2)} 
	\,, 
}
where index $(i)$ denotes the simulated maps obtained from $i$th set of Monte Carlo simulation, 
and $\ave{\cdots}_{(i)}$ denotes the ensemble average for the $i$th set of Monte Carlo. 
Note here that $\ave{n_{\bl}^{ab,cd}}$ corresponds to the disconnected part of 
$\ave{\barx^{ab,(C)}_{\bl}\barx^{cd,(C)}_{-\bl}}$. We then obtain 
the derivative of a log-likelihood, $\mC{L}=\ln P $, as 
\al{
	\PD{\mC{L}}{C^{xx}_\l} &= \frac{1}{P}\PD{P}{C^{xx}_\l} \simeq \frac{1}{P\rom{g}}\PD{P}{C^{xx}_\l} 
	\notag \\ 
	&= \sum_{abcd}\Big[ \barx^{ab,(C)}_{\bl}\barx^{cd,(C)}_{-\bl} - n_{\bl}^{ab,cd} + 
		(a\leftrightarrow c) + (a\leftrightarrow d)\Big] 
	\,. \label{Eq:deriv-P}
}

\subsubsection{Temperature} 

Here we first consider the case of temperature alone (or $a=b=c=d\equiv X$). 
With \eq{Eq:deriv-P}, the derivative of a log-likelihood, $\mC{L}=\ln P $, is 
\al{
	\PD{\mC{L}}{C^{xx}_\l} = 3[|\barx^{(C)}_{\bl}|^2 - n_{\bl}] 
	\,, \label{Eq:deriv-T}
}
where we drop the index, $X$, in the unnormalized estimator and disconnected bias. 
\eq{Eq:deriv-T} motivates an unbiased estimator: 
\al{
	\widehat{C}^{xx}_\l = \left(\frac{A_\l}{2}\right)^{2}[|\barx^{(C)}_{\bl}|^2 - n_{\bl}] 
		= |\estx^{(C)}_{\bl}|^2 - \widehat{n}_{\bl}
	\,, 
}
where 
\al{
	&\estx^{(C)}_{\bl} = \frac{A_\l}{2}\sum_{\bl_1}f_{\bl,\bl_1}\ol{\T}_{\bl_1}\ol{\T}_{\bl-\bl_1} 
	\\ 
	&\widehat{n}_{\bl} = 2[2\ave{\estx^{\T^{(1)}\T}_{\bl}\estx^{\T^{(1)}\T}_{-\bl}}_{(1)}
		- \ave{\estx^{\T^{(1)}\T^{(2)}}_{\bl}\estx^{\T^{(1)}\T^{(2)}}_{-\bl}}_{(1),(2)}] 
	\,. \label{Eq:est-noise}
}
The above equation coincides with \cite{Namikawa:2012pe}, and is generalized for the cases using 
only lensed E-/B-modes alone. 

\subsubsection{Temperature and Polarizations} 

We now generalize the case including polarizations. 
Eq.~(\ref{Eq:deriv-P}) implies that, for each combination of $(a,b)$ and $(c,d)$, we can construct 
the lensing power spectrum estimator as 
\al{
	&\widehat{C}^{ab,cd}_\l = \estx^{ab,(C)}_{\bl}\estx^{cd,(C)}_{-\bl} - \widehat{n}_{\bl}^{ab,cd} 
	\,, 
}
where the first term is the power spectrum of the usual quadratic estimator but the second term is 
the estimator for the disconnected bias, $\widehat{n}^{ab,cd}$, defined as 
\al{
	&\widehat{n}^{ab,cd}_{\bl} = \ave{(\estx^{a^{(1)}b}_{\bl}+\estx^{ab^{(1)}}_{\bl})
		(\estx^{c^{(1)}d}_{-\bl}+\estx^{cd^{(1)}}_{-\bl})}_{(1)}
	\notag \\ 
	&\qquad - \frac{1}{2}\ave{(\estx^{a^{(1)}b^{(2)}}_{\bl}+\estx^{a^{(2)}b^{(1)}}_{\bl})
		(\estx^{c^{(1)}d^{(2)}}_{-\bl}+\estx^{c^{(2)}d^{(1)}}_{-\bl})}_{(1),(2)}
	\,, 
}
Denoting $\alpha,\beta=abcd$, an optimal estimator would be obtained by combining all combinations 
of $abcd$ as 
\al{
	\widehat{C}^{xx}_\l = N_\l^{xx} \sum_{\alpha,\beta}\{(N_\l^{xx})^{-1}\}^{\alpha,\beta} \widehat{C}^\alpha_\l
	\,, 
}
where the optimal noise level and noise covariance matrix are given by 
\al{
	&N_\l^{xx} \equiv \sum_{\beta,\beta'}\{(\bR{N}^{xx}_\l)^{-1}\}_{\beta,\beta'}  \,; \quad 
	\{\bR{N}_\l\}^{\alpha,\beta} = \ave{\widehat{C}^\alpha_\l\widehat{C}^\beta_\l}  \,. 
}

\section{Numerical Simulation of Lensed CMB Maps} \label{app:sim}

In this section, we briefly present our procedure to prepare lensed CMB maps. Our procedure is same as 
for the lensed CMB temperature maps in our previous paper (\citealt{Namikawa:2012pe}, Appendix A), but 
newly including the polarization fluctuations. We prepared lensed CMB maps as follows:

1) We obtain unlensed CMB temperature and polarization power spectra, $C_\l^{\T\T}$, $C_\l^{\T E}$ and 
$C_\l^{EE}$, with CAMB \citep{Lewis:1999bs}. 

2) We generate Gaussian temperature fluctuations $\T_{\bl}$ in Fourier space, based on the input 
power spectrum $C_\l^{\T\T}$. Then, we also generate polarization fluctuations 
$E_{\bl}=\sqrt{C_\l^{EE}-(C_\l^{\T E})^2/C_\l^{\T\T}}R_{\bl}+(C_\l^{\T E}/C_\l^{\T\T}) \T_{\bl}$
where $R_{\bl}$ is normalized Gaussian fields (with zero mean and unit variance). Then, the 
fluctuations of $\T_{\bl}$ and $E_{\bl}$ satisfy the input power spectra. Here we assume 
the primordial B-mode is zero. By performing a Fourier transform on the fluctuations ($\T_{\bl}$ 
and $E_{\bl}$), we generate an unlensed CMB map. The map is a square of $\sqrt{4\pi}$ radian 
($\simeq 203$deg) on a side. We prepare 100 such unlensed maps.

3) We make a lensed CMB map by remapping the unlensed map according to Eq.(1). Here, we perform the 
ray-tracing simulations to obtain the deflection angles. We used a publicly available code RAYTRIX 
(\citealt{Hamana:2001vz}) which follows the multiple scattering. In the standard multiple lens plane 
algorithm, we divide the distance from the observer to the last scattering surface (LSS) into several 
equal intervals and then put lens planes in the every intervals. The light rays emitted from the 
observer are deflected in the every lens planes before reaching the LSS. We numerically solve the 
light-ray positions by solving the multi lens equation and finally obtain the angular position shifts 
on the LSS (See Namikawa et al. (2013), Appendix A, for detailed discussions). We checked that the 
power spectrum of the lensing potential agrees with the expectation from CAMB. 

4) By repeating the procedures 1) to 3), we prepared 100 lensed CMB maps. Each map has an area of 
$10\times 10 {\rm deg}^2$ with $1024\times 1024$ grids, and hence the resulting angular resolution is
$10 {\rm deg}/1024 \simeq 0.6$arcmin. Note that, in our analysis of lensing reconstruction, we further 
cut the maps into $5\times 5 {\rm deg}^2$. 

\begin{figure}
\vspace*{1.0cm}
\includegraphics[width=80mm]{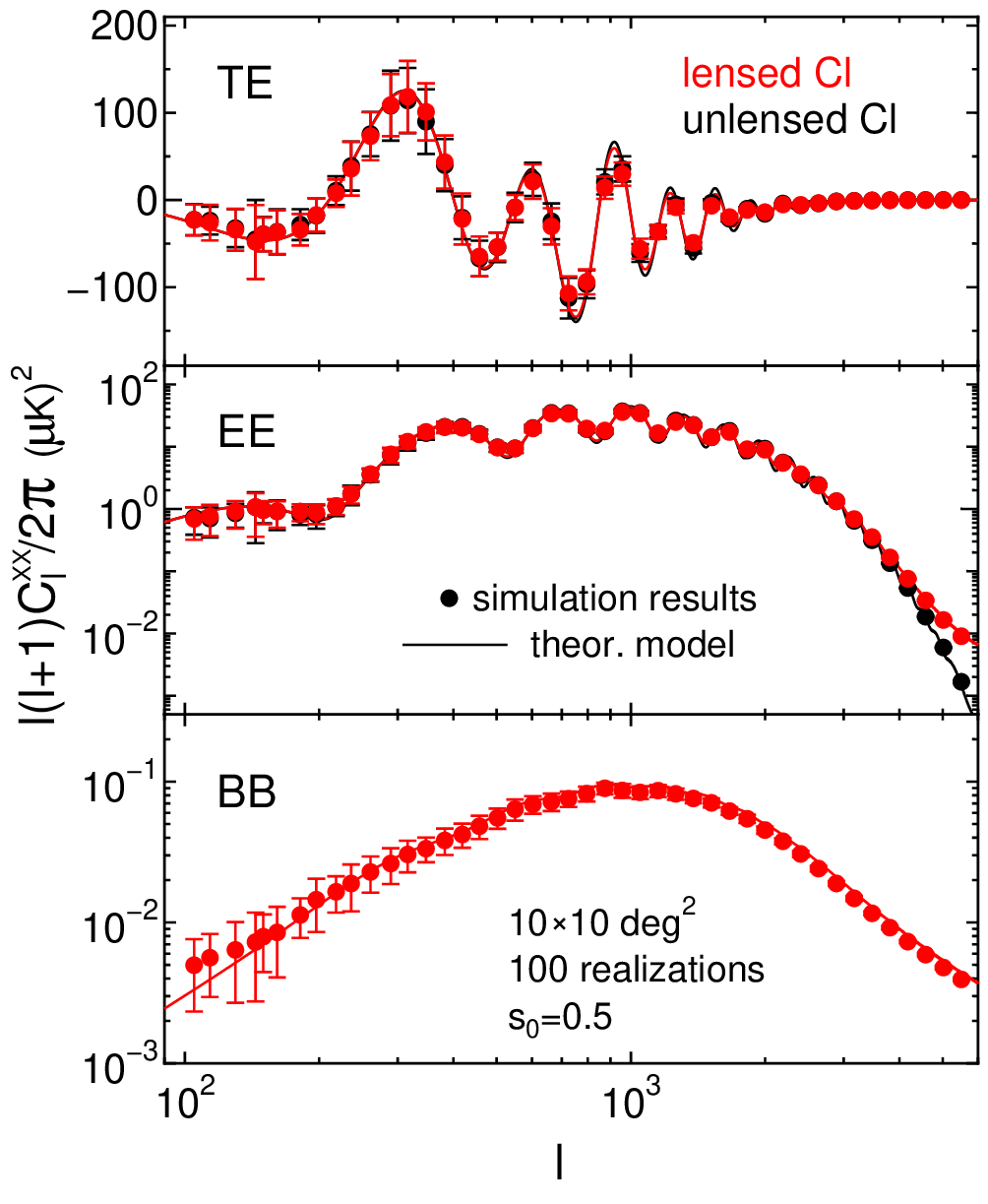}
\caption{
The lensed CMB power spectra for the TE(upper), EE(middle) and BB(lower). The red (black) symbols are 
the lensed (unlensed) power spectrum. The dots with error bars are our simulation results calculated 
from the $100$ realizations of $10 \times 10 {\rm deg}^2$ maps. The solid curves are the theoretical 
prediction of CAMB.
}
\label{fig_appendix_1}
\vspace*{0.5cm}
\end{figure}

Fig.~\ref{fig_appendix_1} shows the CMB power spectra calculated from the $100$ lensed CMB maps.
The upper, middle and lower panels are for the TE, EE and BB power spectrum, respectively. 
The dots with error bars are the mean and the dispersion calculated from the $100$ realizations.
We use $s_0=0.5$ for the apodization given in Eq.~\eqref{apo}. Note that, in order to mitigate the 
effect of E-B mixing due to the survey boundary effect, we estimate the lensed E and B modes using 
``pure'' E and B estimators \citep{Smith:2006vq}. The red(black) symbols are the results for the 
lensed(unlensed) case. The solid curves are the theoretical prediction of CAMB.
Our simulation results agree with the theoretical prediction very well.

\def\eprint#1{preprint (arXiv:#1)}

\bibliographystyle{mn2e-arXiv}

\end{document}